\documentclass{article}

\usepackage{arxiv}

\usepackage[utf8]{inputenc} 
\usepackage[T1]{fontenc}    
\usepackage{hyperref}       
\usepackage{url}            
\usepackage{booktabs}       
\usepackage{amsfonts}       
\usepackage{nicefrac}       
\usepackage{microtype}      
\usepackage{lipsum}		
\usepackage{graphicx}
\usepackage[square,sort,comma,numbers]{natbib}
\usepackage{doi}

\usepackage[T1]{fontenc}
\usepackage{graphicx}
\usepackage{amsmath,amssymb,amsfonts}
\usepackage[ruled,vlined]{algorithm2e}
\usepackage{textcomp}
\usepackage{xcolor}
\usepackage{subcaption}
\usepackage{float}

\usepackage{hyperref}
\usepackage{CJKutf8}
\usepackage{multirow}
\usepackage{graphicx} 
\usepackage{pgfplots}
\usepackage{tikz}
\usepackage{booktabs}
\usepackage{flushend}
\usepackage{amsmath}
\usepackage{amssymb}
\usepackage{adjustbox}
\usepackage{array}

\usepackage{pifont}     

\usepackage{tabularx}
\usepackage{booktabs}
\usepackage{hyperref}
\usepackage{flushend}

\title{Co-Learning: Code Learning for Multi-Agent Reinforcement Collaborative Framework with Conversational Natural Language Interfaces}

\author{
  Jiapeng Yu$^{1,*}$,
  Yuqian Wu$^{1,*}$,
  Yajing Zhan$^{1}$,
  Wenhao Guo$^{1}$,
  Zhou Xu$^{1}$,
  Raymond Lee$^{1,\dagger}$ \\
  \\
  $^1$Beijing Normal University-Hong Kong Baptist University United International College\\ Zhuhai, China \\
  \\
}
\renewcommand{\headeright}{}
\renewcommand{\undertitle}{}
\date{} 
\begin{document}
\maketitle

\renewcommand{\thefootnote}{\fnsymbol{footnote}}
\footnotetext[1]{These authors contributed equally to this work.}
\footnotetext[2]{Corresponding author: \texttt{raymondshtlee@uic.edu.cn}.}

\vspace{-2em}
\begin{abstract}
Online question-and-answer (Q\&A) systems based on the Large Language Model (LLM) have progressively diverged from recreational to professional use. This paper proposed a Multi-Agent framework with environmentally reinforcement learning (E-RL) for code correction called Code Learning (Co-Learning) community, assisting beginners to correct code errors independently. It evaluates the performance of multiple LLMs from an original dataset with 702 error codes, uses it as a reward or punishment criterion for E-RL; Analyzes input error codes by the current agent; selects the appropriate LLM-based agent to achieve optimal error correction accuracy and reduce correction time. Experiment results showed that 3\% improvement in Precision score and 15\% improvement in time cost as compared with no E-RL method respectively. Our source code is available at: \href{https://github.com/yuqian2003/Co_Learning}{https://github.com/yuqian2003/Co\_Learning}.
\end{abstract}
\section{Introduction}
\noindent Large Language Model (LLM) based Conversational Question Answer such as ChatGPT become prominent deep learning networks recognized by daily affairs enquiries to professional tasks solution~\cite{nijkamp2022,barrault2023}. 
They can demonstrate reasoning, planning strengths to match an autonomous agent’s definition to perceive its surroundings, make decisions, operate and even build multi-agents to solve complex problems \cite{xi2023,zhou2023}. Although the majority of multi-agent frameworks can usually complete stationary streaming tasks using fixed prompts but are unable to select the optimal agent according to specific task content \cite{wang2024}.

Program coding involves time-consuming professional skill due to the specific task’s requirement. Beginners often strived for code understandings but may cede programming due to the lack of guidance to resolve unforeseen errors. This paper proposes a \textit{Code-Learning} community based on LLM multi-agent framework on codes correction, annotation for efficient learning in communication with users. It uses reinforcement learning to decide which agent is required for the next step based on the input problem or the output generated by current agent, in contrast to previous multi-agent code generation, error-correction networks based on a defined single stream \cite{chen2023}. 
There are 5 agents responsible for different tasks: 1) Main Agent supervises and exchanges information with users, 2) Correction Agent revises programming, 3) Interpretation Agent explains the programming logic to subsequent agents to locate incorrect codes, 4) Test Agent generates correct codes and 5) Annotation Agent adds comments to the revised code for user’s understanding. 
These five Agents communicate through conversation interfaces. The Multi-Agent generated by the main one is a copy with E-RL to self-improve and feedback to both counterparts and human users. 
Co-Learning uses ERNIE \cite{sun2021}, SparkDesk \cite{iflytek2023}, and LLaMa \cite{touvron2023} as base models for different agents. 
Code error correction with E-RL performance is evaluated by passing probability tests, single loop computation time and numbers of loop required etc. The annotation results are evaluated by an expert reviewer through the location, accuracy, and comprehensibility of annotations.

\noindent The contributions of this paper are to:
\begin{enumerate}
    \item build a Multi-Agent framework based on Multi-LLMs for code error correction.
    \item use original error code datasets to evaluate the performance of multiple LLMs.
    \item explore the possibility of reinforcement learning for large language model based multi-agent operating environment.
    \item compare benchmark frameworks to indicate significant accuracy and operating speed improvements.
\end{enumerate}

\section{Related Work}

\subsection{Prompting with Feedback}
Recent research on large language models has shown that effective use of prompt words can reduce adverse output \cite{gang2023} and induce LLM to generate crucial assessments \cite{wang2024}. 
Prompt engineering is a specialized study with remarkable benefits for reasoning type tasks \cite{peng2024}. 
Reflexion \cite{ahn2023} pointed that using linguistic feedback can reinforce LLM instead of weights to store the feedback text in memory, and induce the large language model to make better decisions, allowing the language agent can learn by mistakes efficiently. 
DEAR \cite{nair2023} can improve LLM judgement in clinical medicine by simulating two agents converse with each other, so that the researcher agent can process information, extract key points of the problem and the decision maker agent integrates them from the researcher agent to judge the final output accordingly. Self-Debugging \cite{chen2023} interprets its self-generated code assisting LLM to identify code errors without explicitly pointing out the errors and modifications by mimicking a rubber duck testing performed by human programmers without extra instructions. When the generated code fails to pass the test, Co-Learning interprets and modify its self-generated code according to the memorized linguistic feedback. At the same time, reinforcement learning will automatically select the optimal agent for the next action based on the feedback from the current agent.

\subsection{Multi-Agent Framework}

\noindent Multi-Agent frameworks emerged at the end of the 20th century \cite{bradshaw1997}, when software engineers used Java to write multi-agents for computers to perform by splitting into small, separate tasks that allow agents to focus and co-operate with each other. At the beginning of the 21st century \cite{bellifemine1999}, JADE \cite{lee2005} standardized the Multi-Agent standard and was used in finance, trading, and journalism \cite{lee2001}.

PADE is Multi-Agent framework based on Python \cite{zanin2019}. LLMs upsurge in GPT \cite{touvron2023} allowing single agents to replace programs \cite{sun2021}. 
LLM’s cognitive abilities in single agents have provided a Multi-Agent foundation \cite{sumers2023}. 
There are many experiments shown that complex, dynamic tasks can be completed by multiple large language model agents equipped with strategies and communications\cite{zhang2023}. 
Hence, Co-Learning uses a PADE framework to create agents with functions, whereas multiple LLMs as the core, component Multi-Agent framework, and information transfer between individual agents to achieve a dynamic workflow.

\subsection{Reinforcement Learning}

Reinforcement learning (RL) is a kind of machine learning where machines interact with an environment to achieve objective \cite{kaelbling1996}. 
During each interaction, the machine considers the environment’s current state, makes decision, operates, observes changes, and transfers feedback into rewards for subsequent round’s state. It aims to maximize the expected cumulative reward over time. The agent that represents the decision-making machine in RL, not only perceives environmental information but also modifies the environment through its actions \cite{lillicrap2015}. 
The agent's perception involves making limited observations of the environment state, such as observing the board situation in Go \cite{silver2016} or the road condition in a self-driving car \cite{kiran2021}, and computes feasible actions by considering the state and policy based on these observations. 

LLMs’ current reinforcement learning mainly focuses on fine-tuning techniques based on feedback \cite{li2023}, minor substitution weights via memory \cite{nair2023, chen2023,wu2024deepfeatureembeddingtabular}, using prompts to help large language models for better strategy decisions. Hence, Co-Learning uses memory while using the current agent's output and performance as reward or penalty to select the best agent for the next action and enable reinforcement learning at environment operating level.

\section{Methodology}

\subsection{Proposed Multi-Agent code correction framework (Co-Learning)}

\begin{figure}[t]
    \centering
    \includegraphics[width=\textwidth]{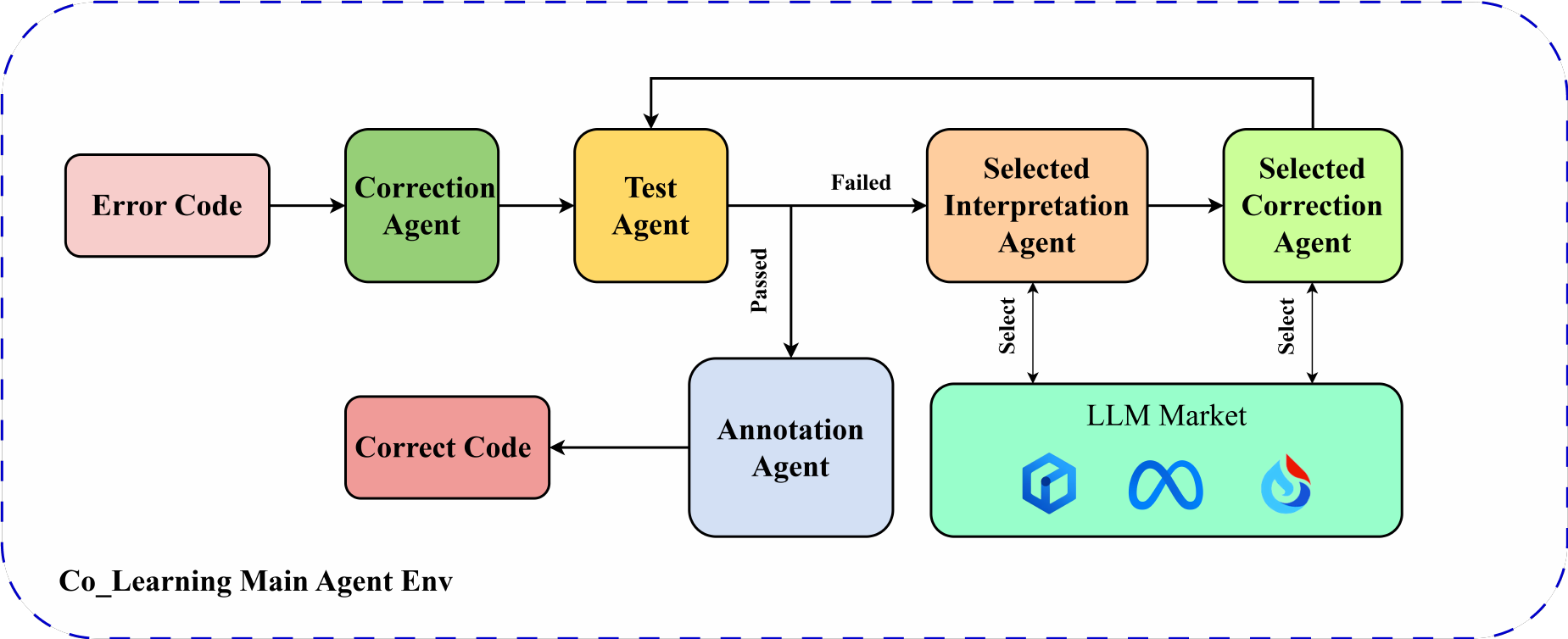}
    \caption{Framework of Multi-Agent code correction (Co-Learning)}
    \label{fig:framework}
\end{figure}

\noindent A Co-Learning Multi-Agent code correction framework is illustrated in Fig. \ref{fig:framework} It has a coexistent framework that relies on PADE \cite{zanin2019} and the entire workflow runs in the environment created by the Main Agent. To begin, a Correction Agent uses a default large language model to make an initial modification for the input error code, returns the generated sentence, and transmits it to the Test Agent. 
The Test Agent performs tests based on the test samples from data set. If the code passes all tests meaning the generation is correct, the Test Agent will send the code to the Annotation Agent for annotation and output it as correct code. If the code is unable to pass any test, the generated code will be passed to the Interpretation Agent and store the interpretation in memory as an environmental reinforcement learning prompt. Then, an error code will transfer to the Correction Agent selected by reinforcement learning to re-generate a code based on the memorized code and interpretation. A loop will be formed by passing the generated result back to the Test Agent. Error codes entered by outsiders during actual use are not included the test cases in the test dataset., three forms of tests used by Test Agent will be provided: test samples entered by the user, test samples generated by LLM based on user-typing-requirement, and the code correctness determined directly by LLM.

For different agents, Main Agent stores all hyper-parameters and historical information, using E-RL based on other agent feedback to update the state of environment. Test Agent create Namespaces to declare the generated code, using test cases to check the generated code, return test results and error messages. Rest of Agents clarify their tasks according to the prompt words, combine with historical information to generate input streams for the LLM, and return results to the Main Agent for storage. Co-Learning involves agents cooperating with each other, mimicking human rubber duck testing while using unit test feedback, selecting the most appropriate large language model in a real-time manner based on E-RL, to enhance the performance of code error correction.

\subsection{Python Agent Development (PADE)}

Python Agent DEvelopment (PADE) is a simple python-based approach to create agents that can be accessed by different devices \cite{zanin2019}. This enables the development and create communication networks for different agents according to FIPA standards. PADE is an architecture based on Twisted to develop a multi-agent application using its library resources (Library) and perform a Running Environment (Running Environment) of a distributed system. PADE controls the platform by creating an agent (Agent Management System) responsible for platform operations, realize for internal platform functions and migrate agents out of the platform to other platforms.

\begin{figure}[t]
    \centering
    \includegraphics[width=0.6\textwidth]{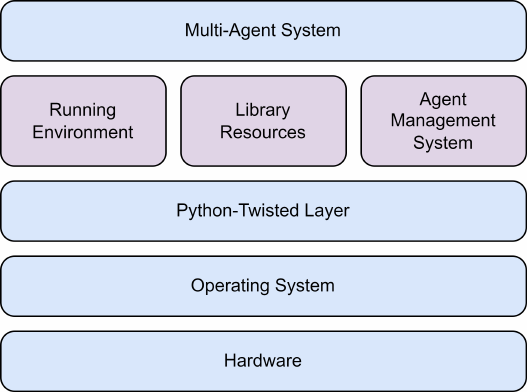}
    \caption{PADE Architecture}
    \label{fig:pade_architecture}
\vspace{-6mm}    
\end{figure}
A PADE architecture is depicted in Fig. \ref{fig:pade_architecture}. It consists of seven modules with the following functions:
\begin{itemize}
    \item \textbf{Core}: All Agents will inherit this base Agent kernel framework when created;
    \item \textbf{Behaviors}: A behavioral template implemented by the Agent that can be inherited by the user to define a variety of personalized behaviors according to FIPA standards;
    \item \textbf{Agent Communication Language (ACL)}: A language model set up for information interaction between agents according to FIPA standards;
    \item \textbf{WEB}: A Web server with graphical interface for interaction with registered sessions, agents and message databases;
    \item \textbf{CLI}: Functions interaction with PADE platform;
    \item \textbf{Miscellaneous (Misc)}: General functions such as looping a standard form of agent initialization message on the screen;
    \item \textbf{Tests}: Module testing;
\end{itemize}

\begin{figure}[H]
    \centering
    \includegraphics[width=0.8\textwidth]{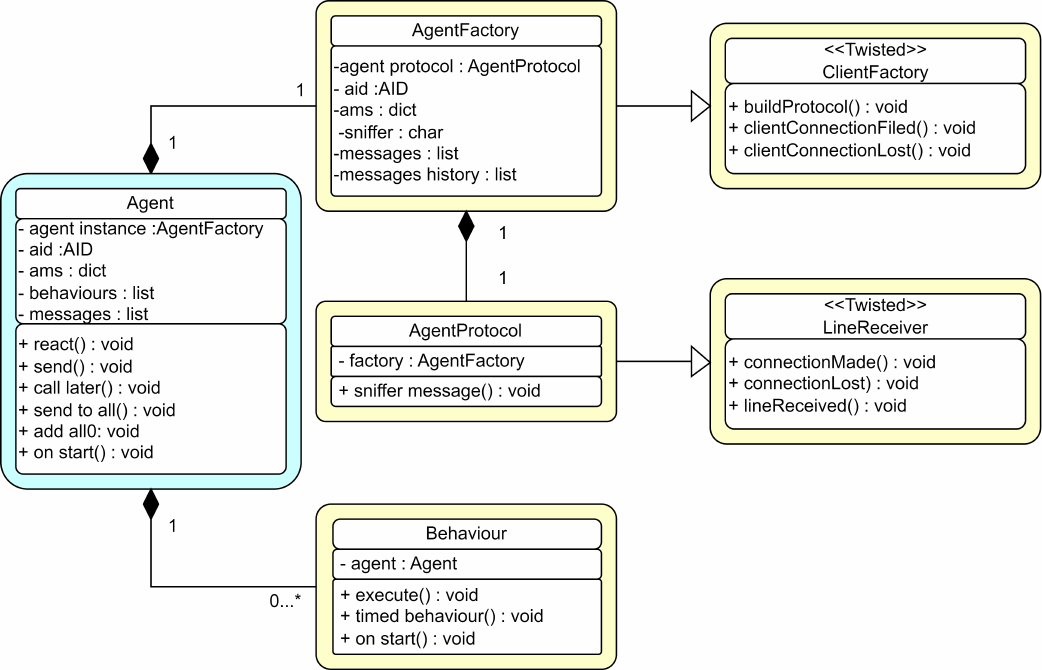}
    \caption{UML Standard for Classes in PADE Framework}
    \label{fig:uml_standard}
\end{figure}
The core of PADE framework is Agent execution. Fig. \ref{fig:uml_standard} illustrates an agent execution UML structure which indicates the class agent and its interaction with other class agents.
\begin{figure}[H]
\centering
\includegraphics[width=0.8\textwidth]{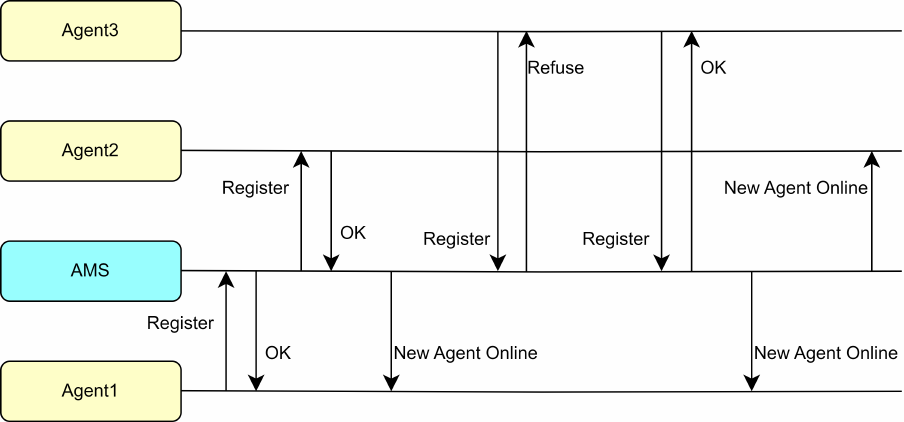}
\captionsetup{belowskip=0pt,aboveskip=0pt}
\caption{Registration and Update of Information in An Example Agent Platform}
\label{fig:agent_interaction}
\vspace{-6mm}
\end{figure}
All agents use PADE framework (e.g., AgentFactory, AgentProtocol in Fig.\ref{fig:uml_standard}) inherit the Agent template and conform to the Twisted protocol. They are identified by own Agent IDentifier (AID). An Agent can be seen is regarded as a connected node in the server platform network that can initiate message exchanges or respond to requests from other network nodes via their AIDs.

\noindent An Agent Management System (AMS) in PADE implements key functions such as control and supervision through a table containing AIDs of all agents according to the FIPA 00023 standards. As an agent, it is the first one to activate followed by the rest of agents are required to register and activate. Each Agent in the network is required to send a message to the AMS agent so that its AID can be saved as a text string by the AMS. Then each Agent in the platform can access to a table in AMS that stores the names and addresses of all Agents to identify by other Agents. The AMS is updated and distributed whenever an agent enters or exits the network.

\begin{figure}[t]
    \centering
    \includegraphics[width=\textwidth]{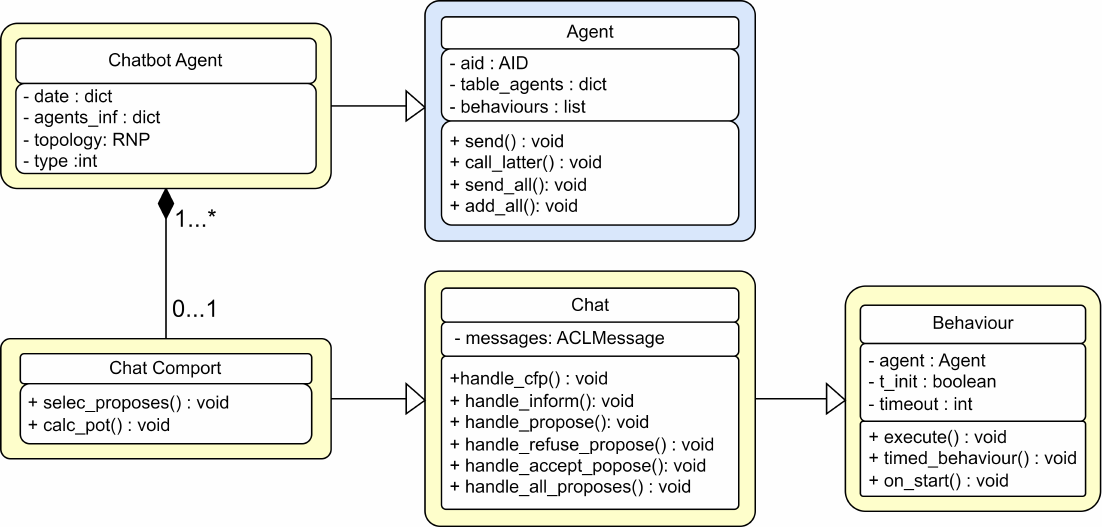}
    \caption{UML Standard for a Chat Agent with FIPA Chat Behaviour}
    \label{fig:chatbot_agent}
    \vspace{-10mm}
\end{figure}

Fig. \ref{fig:chatbot_agent} illustrates shows an example where the AMS is the first agent to be launched. It performs registration for Agents 1, 2, and 3 as they enter, and informs the existing agents when a new agent joins the network. In this case, when Agent 3 fails to register for the first time, its address cannot communicate with other agents. It can only send to the other agents after the second attempt is successful.

Therefore, if an agent requires communication with another agent, it will refer to the address of the target agent in its own table without asking the AMS agent. These agents can communicate even if the AMS agent is deactivated. AMS agent can also record message interactions so that all agents in the network will send a copy of the message received to AMS each time.

AMS is the most significant network agent responsible for agent registration, monitoring, updating active agent tables, logging information exchanged among agents, sending orders to change and deactivate agent’s behaviour.

An agent creation has a well-defined pattern of behavioral classes defined by PADE, protocol classes, and information from the class agent provided by core. Therefore, it is common to have an example in PADE that represents a given agent, and many classes required for behaviors examples representation defined by each agent. 
Fig. \ref{fig:chatbot_agent} illustrates the chat behaviour in PADE. Here, the Chat Agent class inherits all necessary features to execute and identify the Agent to communicate with other counterparts. 
The Chat Port inherits all necessary behaviors from the Chat class according to FIPA standards. Also, the Chat class has the base Behaviour class to implement the base startup with methods

\subsection{Environmentally Reinforcement Learning (E-RL)}

The environmentally reinforcement learning (E-RL) aims to provide a structured environment for LLM-based agent to facilitate its effective execution in code error correction and testing. By using E-RL, LLM-based agent can select actions based on the current state and history to perform code error correction and interpretation. The interactive process starts with a user-supplied code task description and error code. The LLM-based agent then first explains the reason for the code error based on the history and then attempts to correct the erroneous code. The generated code is then executed and compared to two test cases with different levels of difficulty. If the code passes all tests, the interaction is terminated and the final corrected code is returned. If the tests are not passed, it implies that the generated code has errors and requires modification according to the interpretation of the LLM-based agent. This iterative process continues until code is generated that passes all test samples or the maximum number of iterations is reached.

In Co-Learning, a discrete state space has defined as the historical dialog records between the user and agents. These dialog records contain information about the interactions between the user and the assistant, covering aspects such as task descriptions, error hints, and code corrections. The action space, on the other hand, records different actions between the user and the LLM-based agent, such as code corrections and code interpretations. We define two different reward mechanisms at the same time.

The first reward mechanism is used to evaluate the performance of the LLM-based agent during the interaction process. Specifically, if the agent passes the basic test sample, it will be rewarded with 2 points; if it passes the difficult test sample, it will be rewarded with 3 points. Conversely, if the agent fails the test, it will be penalized with 0.5 and 0.2 points, respectively.
\begin{figure}[H]
    \centering
    \includegraphics[width=\textwidth]{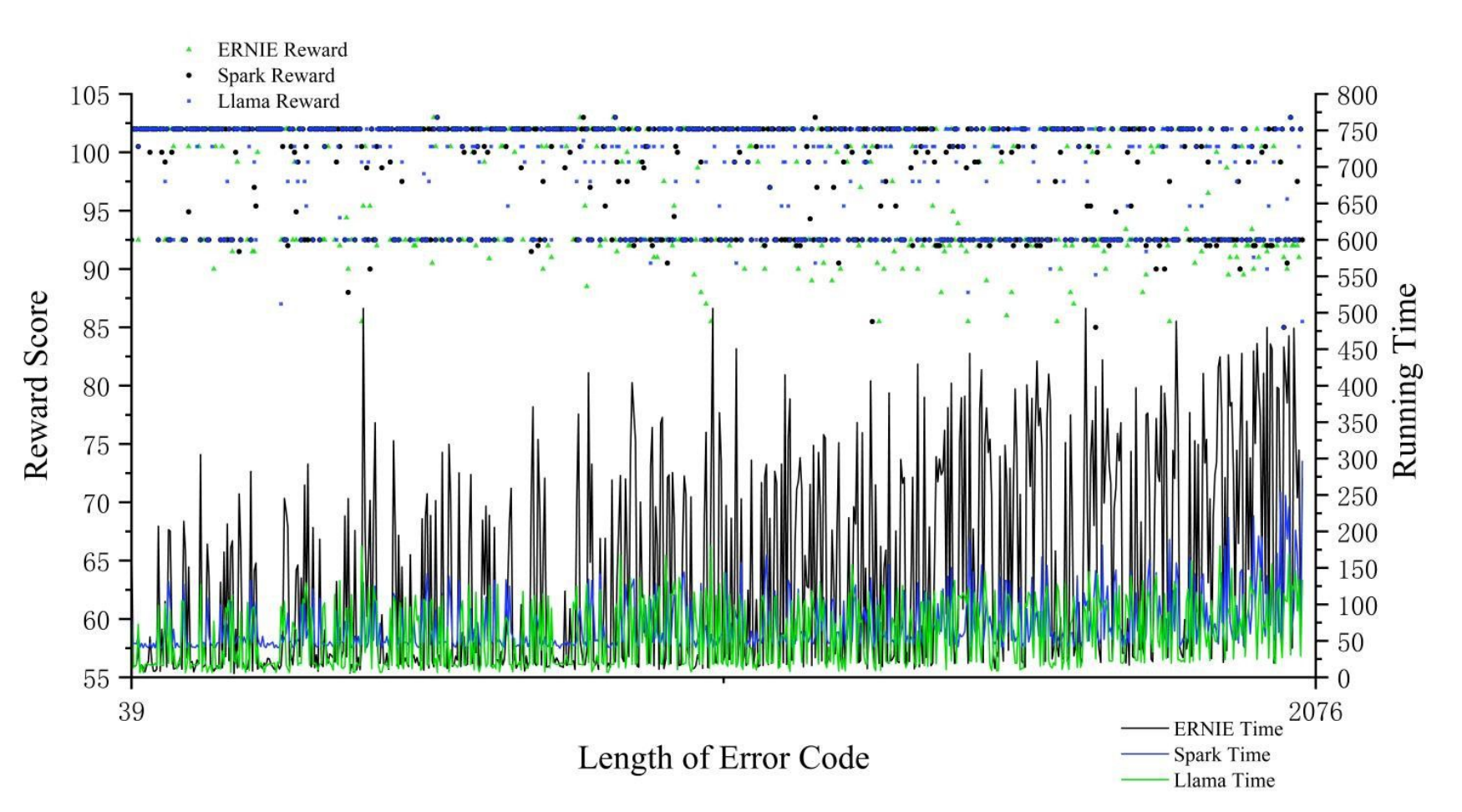}
    \caption{Reward Scores and Running Time for Code Error Correction of Three LLMs}
    \label{fig:reward_scores}
\end{figure}
Obtained the results in Fig. \ref{fig:reward_scores} based on the first reward mechanism, ERNIE performs the best in terms of performance but runs the slowest, LLAMA is rated as medium in terms of performance and runs at average speed, and Spark shows the worst results although it runs the fastest. Based on these different performance characteristics, we dynamically select the initial large model based on the length of the input code. Depending on the code length, we assigned short codes to Spark, medium-length codes to LLAMA, and longer codes to ERNIE, respectively.

The second reward mechanism aims to dynamically select the applicable language model based on the performance of the LLM-based agent under the code error correction dataset. We first use softmax and logistic functions to obtain the weights of the three large language models on the metrics of time and reward, respectively: 

\begin{equation}
    \text{Softmax}(x) = \frac{x_i}{\sum_{j=1}^{3} x_j}
\end{equation}

\begin{equation}
    \text{Logistic}(x) = \frac{1}{1 + e^{-2(x_i - 0.5)}}
\end{equation}

\noindent where $x$ represents reward or time, $x_1$, $x_2$, $x_3$ correspond to ERNIE, LLAMA2-8B, and Spark V3 in terms of reward and time, respectively.

\begin{table}[H]
\centering
\caption{Number of Loops Required for Code Error Correction of Three LLMs}
\begin{tabular}{|c|c|c|c|c|c|}
\hline
LLM & 1 loop & 2 loops & 3 loops & 4 loops & 5 loops \\ \hline
ERNIE 4.0 & 337 & 60 & 26 & 14 & 265 \\ \hline
Llama 3-8b & 317 & 81 & 32 & 21 & 251 \\ \hline
Spark V3 & 319 & 48 & 14 & 4 & 317 \\ \hline
\end{tabular}
\label{tab:llm_loops}
\end{table}

The reward mechanism then calculates a composite score for each language model based on code length, run time, reward value, number of loops (Table. 1), and stability of the language model. These metrics reflect the performance and operational status of the LLM-based agent during the interaction. Then, based on the calculated composite score, the language model with the highest score is selected as the main language model in the current environment. In this way, the reward mechanism is able to automatically adjust the selected language model according to the actual performance of the LLM-based agent in order to improve the efficiency and accuracy of code error correction and interpretation. The algorithm used by E-RL to select LLM is showed in Algorithm ~\ref{ag:code}.

\begin{algorithm}[H]
\SetAlgoLined
\KwResult{LLM, selected LLM}
\KwIn{code, error codes to be changed, run time, number of executed loops}
\KwData{Recent\_LLM, current use of LLM, LLM\_performance, performance parameters LLMs, Parameter\_weight}

\tcc{Calculate the score for each LLM based on the current environmental state}
\ForEach{(model, performance) $\in$ LLM\_performance}{
    len\_weight $\leftarrow$ code\_length $\times$ Parameter\_weight[``length''] $\times$ LLM\_performance[``reward'']\;
    Reward\_weight $\leftarrow$ Parameter\_weight[``reward''] $\times$ LLM\_performance[``reward'']\;
    time\_weight $\leftarrow$ Parameter\_weight[``time''] $\times$ LLM\_performance[``time'']\;
    run\_time\_weight $\leftarrow$ Parameter\_weight[``run\_time''] $\times$ run\_time $\times$ LLM\_performance[``reward'']\;
}

\tcc{Stability penalty if the selected LLM is the same as the current environmental LLM}
\eIf{model = Recent\_LLM}{
    scores[model] $\leftarrow$ (len\_weight + Reward\_weight - time\_weight - run\_time\_weight) $\times$ LLM\_performance[``stability'']\;
}{
    scores[model] $\leftarrow$ len\_weight + Reward\_weight - time\_weight - run\_time\_weight\;
}

\tcc{Update the LLM used by the environment}
LLM $\leftarrow$ max(scores, key = scores.get)\;

\caption{LLM Options Update}
\label{ag:code}
\end{algorithm}

\section{Experiments}
\subsection{Data Description}
The dataset used in the experiments is based on MBPP test set compiled by \cite{austin2021}, which contains numerous prompting words for code generation, automated test cases about python programming problems. A low level LLM are used to generate original error code for Co-Learning, which used prompting words to try to generate code and saved the generated error code as new dataset. The generated code cannot run properly, but contains correct function name and comments about the logic, preserving valid information for subsequent experiments. Generally, this experiment generated an original dataset containing 702 error codes, test samples and challenge test samples as listed in Table \ref{tab:dataset}.

\begin{table}[h]
    \centering
    \caption{Subset of the Dataset Samples}
    \label{tab:dataset}
    \begin{tabular}{|p{0.2\textwidth}|p{0.35\textwidth}|p{0.4\textwidth}|}
        \hline
        \textbf{Error Code} & \textbf{Test List} & \textbf{Challenge Test List} \\ \hline
        \texttt{def remove\_Occ(string, character):} & \texttt{[assert remove\_Occ("hello","l") == "heo",} & \texttt{[assert remove\_Occ("hellololl","l") == "heolol",} \\
        \texttt{... ...} & \texttt{assert remove\_Occ("abcda","a") == "bcd",} & \texttt{assert remove\_Occ("","l") == "",} \\
        & \texttt{assert remove\_Occ("PHP","P") == "H"]} & \texttt{assert remove\_Occ("","l") == ""]}\\ \hline
        \texttt{def is\_woodall(number):} & \texttt{[assert is\_woodall(383) == True,} & \texttt{[assert is\_woodall(32212254719) == True,} \\
        \texttt{... ...} & \texttt{assert is\_woodall(254) == False,} & \texttt{assert is\_woodall(32212254718) == False,} \\
        & \texttt{assert is\_woodall(200) == False]} & \texttt{assert is\_woodall(159) == True]}\\ \hline
    \end{tabular}
\end{table}

\begin{figure}[h]
    \centering
    \includegraphics[width=\textwidth]{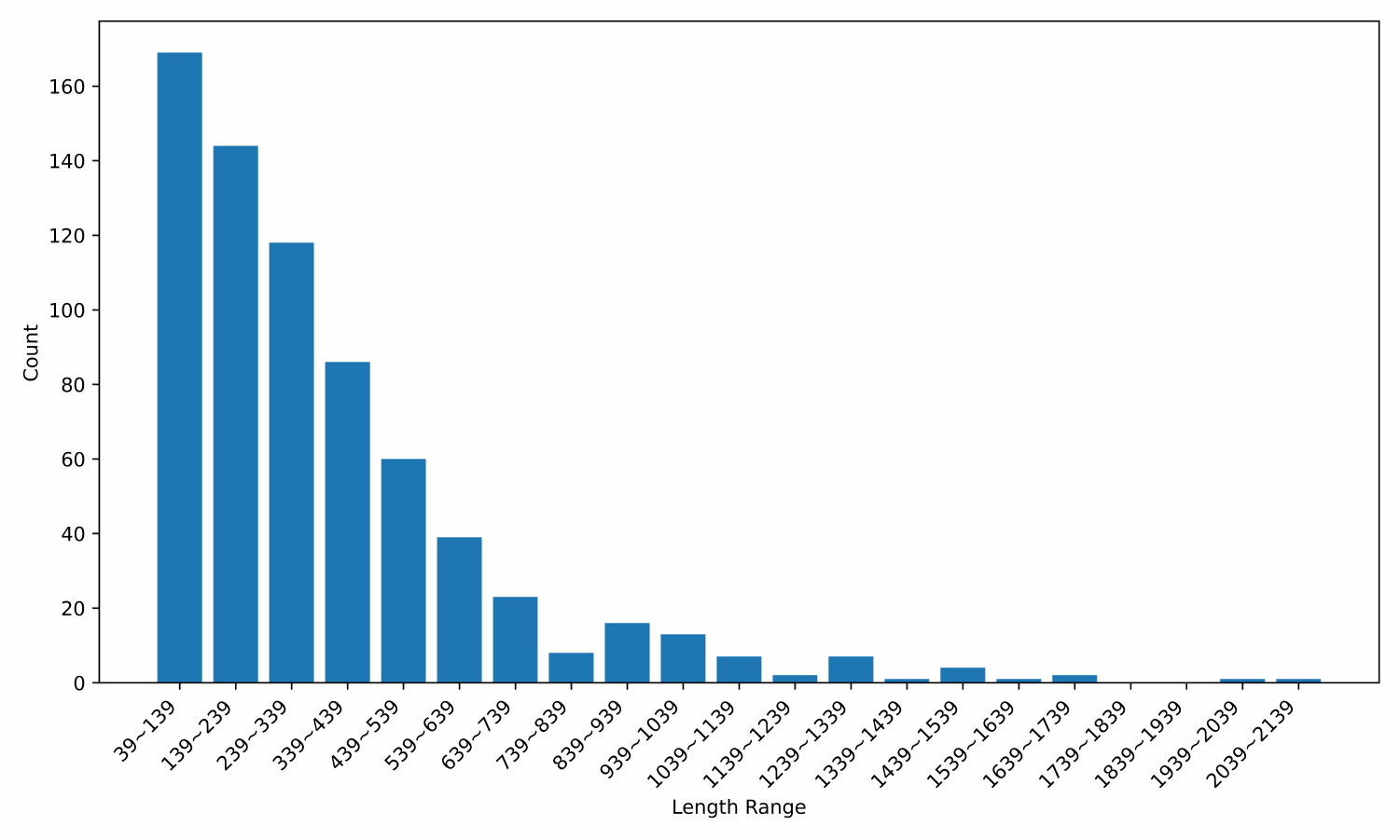}
    \caption{Distribution of Error Code Length Ranges}
    \label{fig:code_length_distribution}
\end{figure}

Fig. \ref{fig:code_length_distribution} shows histograms of dataset error code length distributions. It is instrumental in selecting the best LLM for the environment at the first-time code correction. By categorizing the length of the input message, E-RL will lead Co-Learning to select more capable LLMs for the next agent when obtaining long sentences. An attempt is made to reduce the number of error correction loops while sacrifice a proportion of the generation time, thus reducing the total time consumption and avoiding unnecessary computational costs.

\subsection{Baseline LLM}
ERNIE-4.0-8K-0329 \cite{sun2021}, Spark Desk V3 \cite{iflytek2023} and Meta-Llama-3-8b \cite{touvron2023} are selected as the open-source LLMs. Baseline LLMs are merged into the PADE multi-agent environment. E-RL select the optimal model for the Co-Learning framework from the three to provide high-quality responses.

\subsection{Main Results}

\begin{table}[h]
    \centering
    \caption{Co-Learning with Different LLM Performance Comparison}
    \label{tab:llm_comparison}
    \begin{tabular}{|c|c|c|c|c|c|c|c|}
        \hline
        \textbf{Method} & \textbf{1 loop} & \textbf{2 loops} & \textbf{3 loops} & \textbf{4 loops} & \textbf{5 loops} & \textbf{Average running time (s)} & \textbf{Accuracy (\%)} \\ \hline
        Co-Learning (ERNIE 4.0) & 337 & 60 & 31 & \textbf{29} & 245 & 137.5 & 65.09 \\ \hline
        Co-Learning (Llama 3-8b) & 317 & 81 & 32 & 21 & 251 & 112.8 & 64.24 \\ \hline
        Co-Learning (Spark V3) & 319 & 48 & 14 & 4 & \textbf{317} & 57.7 & 54.84 \\ \hline
        Co-Learning (E-RL)  & 280 & \textbf{104} & \textbf{65} & 27 & 226 & 99.8 & \textbf{67.80} \\ \hline
    \end{tabular}
\end{table}

This experiment uses the original error code data set, sets the maximum number of cycles to 5 (exceeding the number of cycles will directly determine the operation failure), and limits the memory length to 3 dialogue pairs to avoid exceeding the LLM single input message length limit.

Table \ref{tab:llm_comparison} shows the number of successfully corrected loops, average running time, and final accuracy in the code correction task using a single LLM or E-RL and a collaborative learning framework based on multiple LLMs. Co-learning mimicked rubber-duck debugging operations can be observed to help the model retry generation when the first generation goes wrong, with Llama 3-8b being the biggest beneficiary of the single LLM model, with 134 successful re-generations of the correct code.

Through E-RL, the probability of success of Co-Learning based on rubber duck testing is greatly increased, even if some first-time success probability is lost, a total of 196 examples are correctly modified due to E-RL. E-RL's contribution to the runtime is also undeniable, being only slower than the high-speed, low-enabled Spark V3, with an average test time of 99.8s. Finally, Co-Learning with E-RL has the highest review success rate, reaching 67.80\%.

\subsection{Case Study}

\begin{figure}[h]
    \centering
    \includegraphics[width=\textwidth]{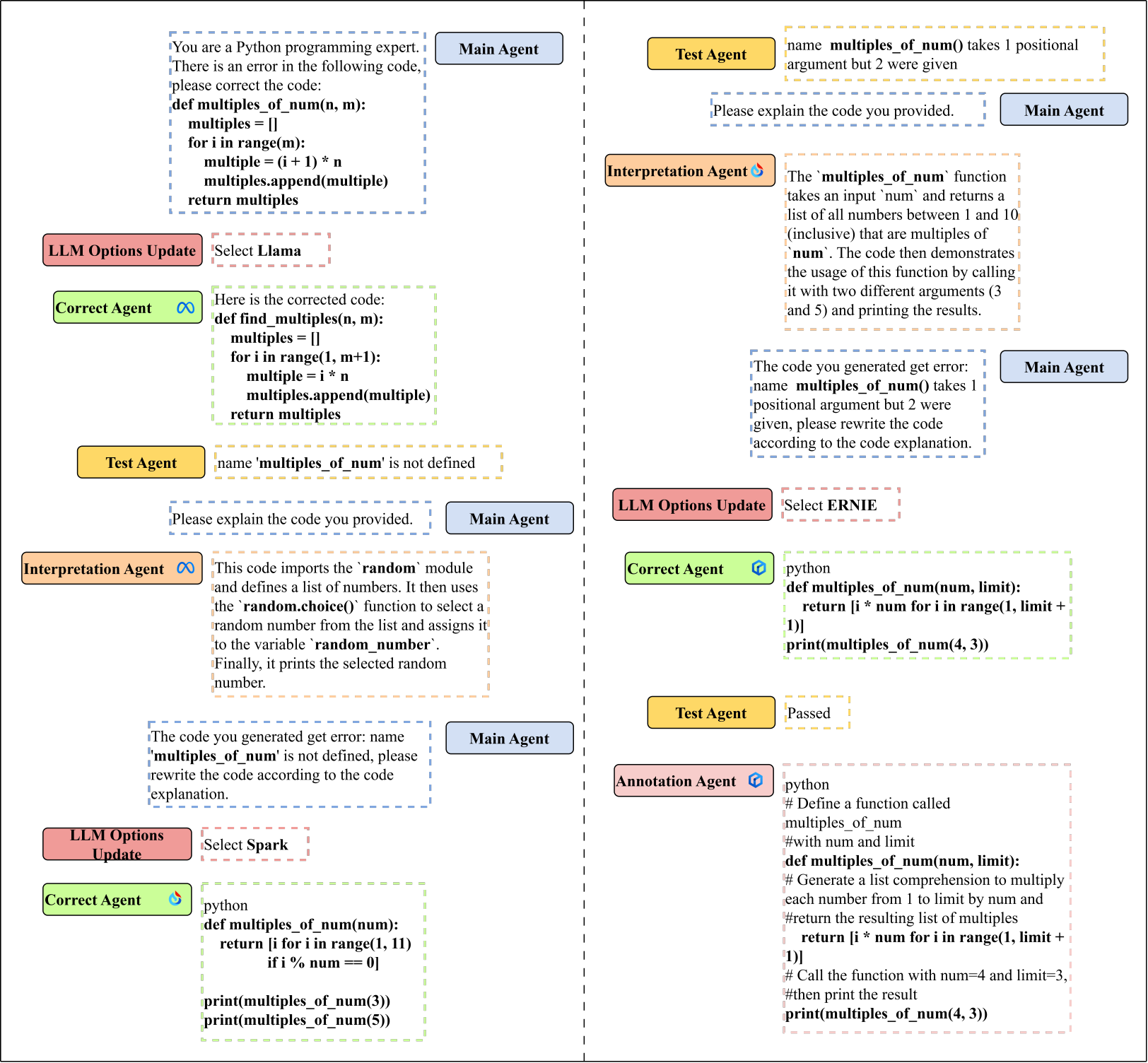}
    \caption{Example of Co-Learning Code Correction}
    \label{fig:co_learning_correction}
\end{figure}

Fig. \ref{fig:co_learning_correction} depicts the actual situation of code error correction through Co-Learning. First, the master agent schedules the error correction task, and E-RL selects Llama as the initial LLM for joint learning based on the input message. Correct Agent uses Llama to make initial modifications to error codes. Based on the results it can be concluded that Llama generated the correct answer but secretly changed the function name. This phenomenon is obvious in the three LLMs. The LLM may choose more appropriate function names for the code based on the code content, while ignoring the user's needs.

The Test Agent detects that the generated code fails the test and returns an error message to the Main Agent. The Main Agent assigns code interpretation tasks in the hope of simulating rubber duck testing. The Interpretation Agent interprets the generated code and stores the contents in the Main Agent memory. E-RL re-selects Spark as the next Agent’s LLM, and the Correct Agent then re-corrects the code. Spark misinterprets the code as outputting multiples of the current number between 1 and 10, which may be related to the fact that the error returned by the Test Agent contains information about the first test sample, where the maximum number to be generated is 10.

Test Agent detected the error, leading Main Agent decide to make another rubber duck debugging. Then LLM was changed to the most powerful ERNIE, which eventually generated the correct code, and Annotation Agent added comments to the output.

This case shows how Co-Learning can continuously correct, understand and then correct erroneous code by imitating the error correction process of human programmers, and generate information-intensive integrated code that only senior programmers can generate. With E-RL, Co-Learning attempts to balance model power and speed in the hope of generating the best response in the shortest possible time, creating code that is shorter and more refined than expected correct code, while using less time than a single model.

\section{Conclusion and Future Works}

This paper focuses on developing a code learning community (aka Co-Learning) framework based on an LLM-based multi-agent framework that leverages ambient reinforcement learning (E-RL) for agent self-improvement. The community aims to interpret error codes and perform code correction tasks to provide users with a more intelligent and personalized programming learning experience. Experiments show that the Co-Learning framework can effectively improve the code error correction capabilities of current LLM. E-RL dynamically determines the state of the environment and changes the selection of the LLM, which can speed up the code correction process and achieve significant improvements in the quality of the generated output.
In the future, Co-Learning will focus on further optimizing the E-RL algorithm to improve the Agent's learning efficiency and performance. At present, it seems too simplistic to select E-RL parameters based only on model capabilities. Making Co-Learning's environmental reinforcement learning have dynamic self-updating weights by combining machine learning will be one of the main goals in the future. Expect to explore more complex tasks and scenarios, including error correction and code understanding in larger code bases, as well as code learning in different programming languages and domains.

\section{Data Availability Statement}
The datasets analyzed for this paper can be accessed upon request from interested readers at \href{https://github.com/yuqian2003/Co_Learning}{https://github.com/yuqian2003/Co\_Learning}.
\section{Conflict of Interest}
The authors declare that the research was conducted in the absence of any commercial or financial relationships that could be construed as a potential conflict of interest.
\section{Author Contributions}
JY: Conceptualization, Methodology, Multi-Agent framework design, Data processing, Writing-original draft preparation and Formal analysis; YW: Data validation, Environmentally reinforcement learning implement, Multi-Agent framework design, Writing-original draft preparation and Formal analysis; YZ, WG and ZX: Literature review, PADE Feasibility Analysis and Data Visualization; RL: Supervision, Reviewing and Editing.

\vspace{12pt}

\end{document}